\pdfoutput=1

\documentclass[11pt]{article}

\usepackage[preprint]{acl}

\usepackage{times}
\usepackage{latexsym}

\usepackage[T1]{fontenc}

\usepackage[utf8]{inputenc}

\usepackage{microtype}

\usepackage{inconsolata}

\usepackage{graphicx}
\usepackage{algorithm}
\usepackage{algorithmic}
\usepackage{multicol, multirow}
\usepackage{enumitem} 
\usepackage{xcolor}
\usepackage{colortbl}
\usepackage{graphicx}
\usepackage{epstopdf}
\usepackage{svg}
\usepackage{arydshln}
\usepackage{dashrule}
\usepackage{tablefootnote}
\usepackage{amsmath}
\usepackage{booktabs}
\newcommand{\changed}[1]{{#1}}
\title{UniSLU: Unified Spoken Language Understanding from \\ Heterogeneous Cross-Task Datasets}

\author{Zhichao Sheng, Shilin Zhou, Chen Gong, Zhenghua Li \\
  Institute of Artificial Intelligence, School of Computer Science and Technology, \\
  Soochow University, Suzhou, China \\
  \texttt{20245227108@stu.suda.edu.cn; slzhou.cs@outlook.com} \\
  \texttt{gongchen18@suda.edu.cn; zhli13@suda.edu.cn}
}

\begin{document}
\maketitle
\begin{abstract}
Spoken Language Understanding (SLU) plays a crucial role in speech-centric multimedia applications, enabling machines to comprehend spoken language in scenarios such as meetings, interviews, and customer service interactions. SLU encompasses multiple tasks, including Automatic Speech Recognition (ASR), spoken Named Entity Recognition (NER), and spoken Sentiment Analysis (SA). However, existing methods often rely on separate model architectures for individual tasks such as spoken NER and SA, which increases system complexity, limits cross-task interaction, and fails to fully exploit heterogeneous datasets available across tasks.
To address these limitations, we propose UniSLU, a unified framework that jointly models multiple SLU tasks within a single architecture.
Specifically, we propose a unified representation for diverse SLU tasks, enabling full utilization of heterogeneous datasets across multiple tasks. Built upon this representation, we propose a unified generative method that jointly models ASR, spoken NER, and SA tasks, enhancing task interactions and enabling seamless integration with large language models to harness their powerful generative capabilities. Extensive experiments on public SLU datasets demonstrate the effectiveness of our approach, achieving superior SLU performance compared to several benchmark methods, making it well-suited for real-world speech-based multimedia scenarios. We will release all code and models at \url{github} to facilitate future research.
\end{abstract}
\section{Introduction} 

\begin{figure}[t]
  \centering
  \includegraphics[width=\linewidth]{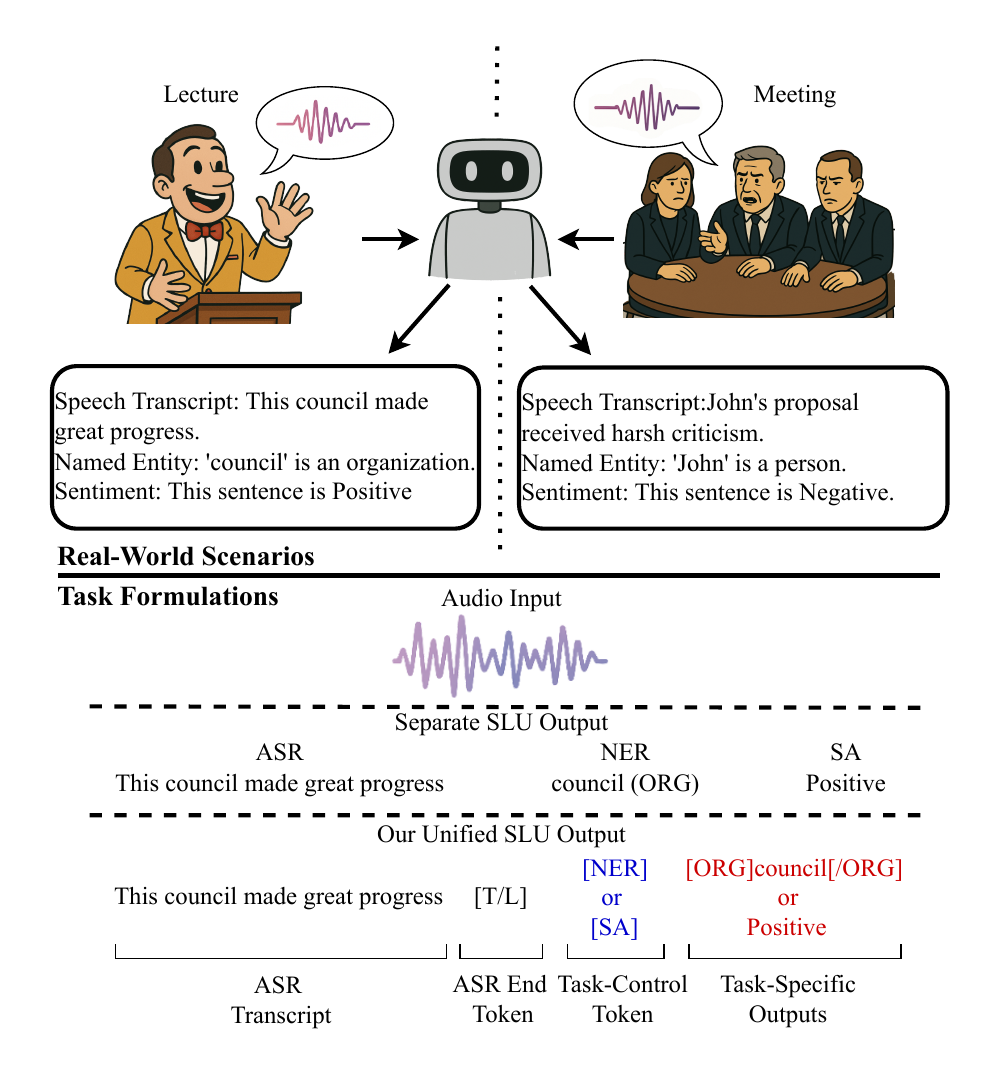}
  \caption{Illustration of real-world SLU application scenarios and corresponding task formulations.}
  \label{fig:data_representation}
\end{figure}

Spoken Language Understanding (SLU) \cite{lakhotia-etal-2021-generative} aims to enable machines to comprehend human speech by converting audio signals into structured, interpretable information \cite{DBLP:conf/mm/WangZ0ZLW0Z24}. As a core component of multimedia systems, SLU bridges the gap between natural speech and machine understanding, supporting a wide range of real-world applications \cite{yuan-etal-2024-openvna} in scenarios such as lectures, meetings \cite{DBLP:conf/emnlp/DuZZY22, luz-etal-2006-gathering}, daily chats \cite{chiba-higashinaka-2025-investigating}, and customer service interactions, as shown in the upper part of Figure \ref{fig:data_representation}.
SLU typically involves multiple tasks. In this work, we focus on three representative tasks: automatic speech recognition (ASR) \cite{imai-etal-2025-evaluating}, named entity recognition (NER) \cite{zhou-etal-2024-chinese}, and sentiment analysis (SA) \cite{DBLP:conf/emnlp/MitsuiMWHS24}. As shown in Figure \ref{fig:data_representation}, ASR transcribes speech into text. Spoken NER identifies named entities—such as persons, organizations, and locations—in the speech, while spoken SA analyzes the sentiment expressed in the speech.
Existing SLU research typically handles different SLU tasks with separate models, designing task-specific modules for each task \cite{ xu2021topic, DBLP:conf/aaai/ZouZKLPJS0HL21, DBLP:conf/iccai/DaiZC23}.
For instance, in the case of spoken NER and SA tasks, SLUE \cite{shon2022slue} employs a separate model approach, adding a classification module specifically for the SA task.
While this approach offers a solution for handling SLU tasks, it still faces several challenges \cite{poria2017review, ghannay2018end}. First, training multiple SLU tasks with distinct models overlooks the potential for interactions across tasks, which could benefit the overall SLU performance. Second, each model is usually trained on task-specific datasets, failing to fully leverage multi-source heterogeneous datasets across tasks, which can result in data sparsity. Lastly, modeling different SLU tasks with separate architectures increases the complexity of the overall model design.



To enhance interactions between multiple SLU tasks and address them in a unified manner, \citet{DBLP:conf/emnlp/KumarMZTVBMSG24} propose a unified model capable of jointly modeling several SLU tasks, including ASR, speaker change \cite{DBLP:conf/acl/ChengZZWCC23}, endpointing, and NER. This model facilitates interactions between tasks, enabling them to share and enrich semantic understanding. However, their approach requires aligned datasets with annotations for all these SLU tasks simultaneously, which are extremely rare and costly to obtain, particularly for the specific tasks we focus on, i.e., ASR, NER, and SA.

To address the above limitations, we propose \textbf{UniSLU}, a unified generative framework that jointly performs diverse SLU tasks within a single model architecture, promoting cross-task interactions and enables effective use of heterogeneous datasets across tasks.
Specifically, we first propose formulating different SLU tasks into a unified representation, enabling full utilization of heterogeneous datasets across tasks. As shown in the bottom part of Figure \ref{fig:data_representation}, the unified representation consists of four components: ``[ASR Transcript]'' which represents the transcribed speech, ``[ASR End Token]'' (i.e., ``[T/L]''), which marks the end of the transcript, ``[Task-Control Token]'', which specifies the downstream SLU task (i.e., ``[NER]'' or ``[SA]''), and ``[Task-Specific Outputs]'', which denote the corresponding task predictions.
Based on the representation, we propose a unified generative framework to jointly perform ASR, spoken NER, and SA tasks, which, on the one hand, enhances interactions across tasks and, on the other hand, can be easily integrated with large language models (LLMs) to leverage their strong generative capabilities.
Extensive experiments on heterogeneous datasets demonstrate the effectiveness of our approach, resulting in overall improvements in SLU performance.

The contributions of our work can be summarized as follows: 
\begin{itemize}
    \item We propose a unified representation for diverse SLU tasks, enabling the model to fully leverage heterogeneous data across different tasks.
    \item We propose a generative framework that jointly models multiple SLU task, unifying them within a single architecture and enhancing multitask interaction through shared knowledge.
    \item We conduct extensive experiments and provide a detailed analysis on the SLUE-VoxPopuli and SLUE-VoxCeleb datasets. Experimental results demonstrate that the proposed UniSLU 
    consistently outperforms strong baselines and achieves superior overall SLU performance, verifying its effectiveness.
\end{itemize}

To promote reproducibility and future work, we will release all code and models on github.

\section{Related Work}\label{sec:ref}




\subsection{SLU Joint Modeling}
{\changed SLU involves processing and comprehending human speech signals, covering core tasks such as ASR, NER, and SA. In this work, we focus on these three key tasks as representative aspects of SLU.
In this section, we briefly review prior work on joint modeling of ASR, NER, and SA, with a focus on how these tasks are integrated.
}

Joint ASR-NER modeling is a highly active research area in SLU systems. Previous studies \cite{yadav2020end, chen2022aishell} have commonly adopted an approach, where NER is implemented by inserting predefined specific-tags \cite{ghannay2018end, yadav2020end}, which explicitly indicate NER boundaries and categories within the ASR transcript. This method integrates the two tasks without requiring an additional module for NER, simplifying the overall model design. However, this approach has certain limitations. While it can handle both ASR and NER information simultaneously, inserting specific tags directly into the ASR transcript may introduce ambiguity in the model's understanding of ASR itself. 
{\changed More specifically, this tag-specific approach in NER departs from natural language expression and lacks flexibility for broader SLU tasks. It reveals a key limitation—task-specific tagging schemes are often difficult to unify with other task formulations, posing challenges for unified modeling across diverse SLU tasks.}

Some recent studies have explored joint ASR and SA modeling. Traditionally, ASR and SA models are trained separately, where the ASR output serves as input for SA or directly outputs the SA output \cite{bertero2016real}. In recent approaches, a shared encoder is used for both ASR and SA, but each task has its own decoder \cite{lu2020speech}. This design allows both tasks to share audio features and enables optimization toward common speech representations during training. However, it also has certain limitations, as it requires a separate module specifically designed for the SA task, increasing model complexity. Additionally, the interaction between tasks remains limited to shared audio features, without deeper semantic integration.


SLUE \cite{shon2022slue} is a widely used benchmark for spoken NER and SA, where spoken NER employs a tag-specific approach, and spoken SA uses a classification module after the audio encoder. These tasks are trained separately, limiting their interaction. The tag-specific method for NER \cite{yadav2020end, ghannay2018end} and the additional SA module do not fully exploit potential cross-task benefits. To address this, \citet{DBLP:conf/emnlp/KumarMZTVBMSG24} proposed a unified model that jointly models multiple SLU tasks, enabling deeper interaction and semantic collaboration. However, this approach faces a major challenge: open-source datasets with unified annotations are scarce and expensive to obtain.

Existing joint methods generally face the following challenges:
First, limited task interaction is a significant challenge in current systems. Although ASR and NER share a common encoder, the lack of true interaction between tasks restricts the potential performance gains that could arise from leveraging the synergy between them. Second, task weight balancing is another issue. Different tasks, such as ASR, NER, and SA, may require varying levels of attention, and balancing the importance of each task during training remains a difficult challenge. Third, data misalignment is a critical concern. In previous work, ASR, NER, and SA often use different data representations, and effectively integrating these representations into a unified framework remains a key research problem. Alternatively, some approaches rely on multi-task aligned labeled data, but this comes with the challenges of high cost and limited availability.

Our work addresses these challenges by introducing a unified data representation and modeling approach, which enhances task interaction and reduces redundant generation, leading to better overall performance. At the same time, it reduces reliance on costly and scarce data by enabling more effective use of heterogeneous and multimodal information, significantly improving the adaptability of SLU systems to real-world multimedia scenarios.
\subsection{Generative Speech Language Models}

Generative speech language models like T5 \cite{raffel2020exploring} and SpeechT5 \cite{ao2021speecht5} extend the ``text-to-text'' framework to speech, unifying tasks like recognition and translation. But, they struggle with SLU tasks like NER and SA due to fundamental differences between speech and text.

Whisper \cite{radford2023robust}, a state-of-the-art speech model, excels in translation and transcription but struggles with NER and SA. Unlike T5, which supports diverse tasks, Whisper is built for speech-to-text and translation, limiting its adaptability to broader SLU tasks.

Recent advancements like FLAN-T5 \cite{chung2024scaling} have aimed to improve multitasking, aligning task descriptions with outputs, but these models primarily focus on text and need further adaptation for speech tasks.

\citet{yang2023chinese} propose a generative multitask approach using Whisper for ASR and NER, adopting a tag-specific method for NER. However, it focuses solely on ASR and NER, limiting task interaction and broader applicability.
In our approach, we integrate ASR, NER, and SA within a unified framework by leveraging Whisper’s encoder for speech understanding and using its decoder to jointly generate outputs for all three tasks. This unified design facilitates cross-task interaction, reduces redundancy, and effectively mitigates challenges such as data misalignment and task imbalance, ultimately improving overall performance. Additionally, by leveraging Whisper’s large-scale pretraining, our model adapts to new SLU tasks with limited additional data, without the need for large task-specific corpora.

\section{Our Approach}\label{sec:approach}
The key idea of this work is to unify multiple SLU tasks—including ASR, NER, and SA—within a unified generative framework. This unified approach is designed to enhance overall SLU performance by learning shared representations and fostering cross-task interaction across heterogeneous datasets. As a result, it is well-suited for complex, multimedia-driven scenarios where spoken content understanding is essential.
In this section, we first present a unified representation for diverse SLU tasks derived from heterogeneous data, and then introduce our proposed generative framework for Unified SLU (UniSLU).
\subsection{Unified Representation of Heterogeneous Data across SLU Tasks}
In previous works, spoken NER and SA are typically addressed by training separate models \cite{shon2022slue}, which requires designing different architectures for each task and limits the interaction of shared knowledge across tasks. \citet{DBLP:conf/emnlp/KumarMZTVBMSG24} enhance multitask interaction by proposing a unified model framework to jointly model various SLU tasks, such as ASR, speaker change detection, endpointing, and NER, relying on aligned data where multiple task annotations are available for each instance. However, such aligned multitask annotation data is scarce, particularly for ASR, NER, and SA.
In contrast, non-aligned heterogeneous datasets for different SLU tasks are more common, where different instances may contain only a subset of the task annotations. For instance, one dataset may provide ASR and NER labels, while another may include ASR and SA annotations, but neither dataset fully covers all three tasks. 

To fully leverage the heterogeneous datasets across tasks, we propose a unified data representation by framing ASR, NER, and SA as a single generative task within a unified template: ``[ASR Transcript][ASR End Token][Task-Control Token][Task-Specific Outputs]''.
In this template,  ``[ASR Transcript]'' refers to the ASR transcript, ``[ASR End Token]'' (we represented as ``[T/L]'') acts as a delimiter separating the transcript from the subsequent task results, ``[Task-Control Token]'' refers to the name of the task (in our case either ``[NER]'' or ``[SA]''), and ``[Task-Specific Outputs]'' represents the output specific to that task. 

For example, in the lower part of Figure ~\ref{fig:data_representation}, the sentence ``This council made great progress'' would be represented for the NER task as ``This council made great progress[T/L][NER][ORG]council[/ORG]'', and for the SA task as ``This council made great progress[T/L][SA]Posi- tive''. This means that ``council'' is identified as an organization entity, and the sentence expresses a positive sentiment.

\subsection{Proposed Unified Generative Framework for Joint SLU} \label{framework}

\begin{figure}[t]
  \centering
  \includegraphics[width=\linewidth]{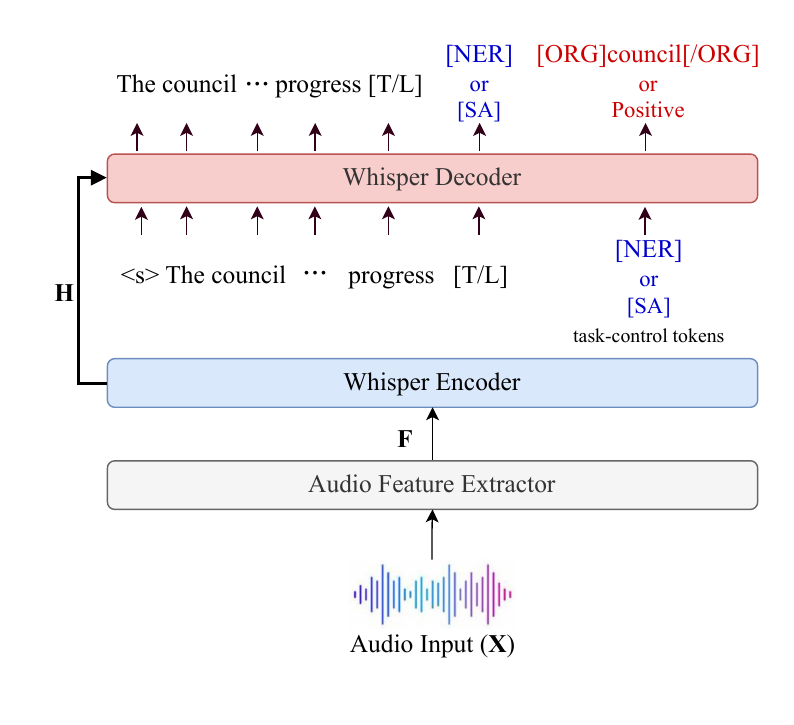}
  \caption{The architecture of our proposed UniSLU framework.}
  \label{fig:whisper}
\end{figure}

In this work, we propose a generative framework to jointly address different SLU tasks, i.e., ASR, NER, and SA, within a unified model, named UniSLU.
Figure~\ref{fig:whisper} illustrates the architecture of UniSLU. The audio feature extractor first extracts raw audio features, followed by the encoder, which captures a more compact, contextual representation. Finally, the decoder generates the ASR transcript and SLU task outputs within a unified sequence.


During generation, the model generates tokens sequentially, starting with the ASR transcript. Once the ``[T/L]'' delimiter is generated, it indicates that the transcript generation is complete, and the model transitions to task-specific outputs. At this stage, we insert the task-control tokens (i.e., ``[NER]'' for the NER task and ``[SA]'' for the SA task) to control the decoding process. When the task-control token ``[NER]'' is inserted, the model is guided to identify the entities and their corresponding types, such as persons, locations, or organizations, based on the previously generated ASR transcript. When the task-control token ``[SA]'' is inserted, the model classifies the sentiment (including Positive, Negative, Neutral, and Mixed) of the spoken content. 
With these inserted control tokens, the decoder is directed to generate subsequent task-specific outputs by leveraging the previously generated sequence, ensuring coherent and context-aware predictions.

In this way, we can facilitate different SLU tasks in a unified generative manner, taking full advantage of shared multitask interactions and leveraging heterogeneous datasets across tasks. Furthermore, the proposed framework can be easily integrated with LLMs to benefit from their strong generative capabilities, as will be detailed in Section \ref{LLM}.

In the following, we provide a detailed description of the model architecture and training strategy.

\paragraph{\textbf{Audio Feature Extractor.}}
The input to the model is a raw speech waveform, denoted as \( \mathbf{X} \), which is first processed into a sequence of feature vectors. Following the preprocessing pipeline of Whisper \cite{radford2023robust}, the audio signal is transformed into a log-mel spectrogram representation to extract audio features before being fed into the encoder. Formally, given an input speech signal \( \mathbf{X} \) of length \( t \), the audio feature extractor produces:
\begin{equation}
\mathbf{F} = \text{MelSpec}(\mathbf{X}), \quad \mathbf{F} \in \mathbf{R}^{t' \times d}
\end{equation}
where \( t' \) is the downsampled sequence length, and \( d \) is the feature dimension.

\paragraph{\textbf{Encoder.}} 
We adopt Whisper's convolutional and Transformer-based encoder due to its strong speech representation capabilities. The encoder consists of a convolutional front-end, followed by stacked Transformer layers, which process the log-mel spectrogram into a sequence of contextualized embeddings. Given the extracted feature sequence \( \mathbf{F} \), the encoder produces:
\begin{equation}
\mathbf{H} = \text{Encoder}(\mathbf{F}), \quad \mathbf{H} \in \mathbf{R}^{t' \times d'}
\end{equation}
where \( d' \) is the encoder hidden dimension. The Transformer layers apply self-attention mechanisms to refine speech features, encoding long-range dependencies within the audio sequence.

\paragraph{\textbf{Decoder.}}
The decoder generates text sequences autoregressively, modeling both ASR transcriptions and SLU outputs within a unified framework. The decoder receives the encoder's output \( \mathbf{H} \) and iteratively generates tokens in a left-to-right manner, conditioned on previous predictions:
\begin{equation}
\mathbf{Y} = \text{Decoder}(\mathbf{H}, \mathbf{Y}_{<t})
\end{equation}
Here, \( \mathbf{Y} \) represents the output token sequence, conditioned on the previously generated tokens \( \mathbf{Y}_{<t} \). The output sequence uses our unified data representation, where ASR transcript is followed by task-specific annotations, such as NER or SA.





\paragraph{\textbf{Dynamic Weighted Training Loss.}} In multitask fine-tuning, balancing the weights between different tasks is crucial. In our case, the ASR transcript text is typically much longer than the NER and SA outputs. Without adjusting the weights, there is a clear tendency for the model to focus more on optimizing ASR, leading to an imbalance in the training process. To achieve this, we design weight assignment strategies for the original cross-entropy loss, dynamic loss based on the proportion of each task's sequence length relative to the full sequence. The basic idea is that longer outputs receive lower loss weights, mitigating the imbalance caused by significantly longer ASR transcript. This ensures that tasks are balanced during training. The total loss function is defined as follows:
\begin{equation}
\mathcal{L}_{total} = W_{\text{task}_i} \mathcal{L}_{\text{task}_i} + W_{\text{ASR}} \mathcal{L}_{\text{ASR}}
\end{equation}
where  $\mathcal{L}_{\text{ASR}}$ and $\mathcal{L}_{\text{task}_i}$ denote the loss functions for ASR and other SLU tasks, both computed using cross-entropy loss, and the corresponding weights  $W_{\text{task}_i}$  and  $W_{\text{ASR}}$  are computed as:
\begin{equation}
W_{\text{task}_i} = \frac{Len_{\text{ASR}}}{Len_{\text{total}}}, \quad
W_{\text{ASR}} = \frac{Len_{\text{task}_i}}{Len_{\text{total}}}
\end{equation}
{\changed where  $Len_{\text{ASR}}$  represents the length of the ASR transcript,  $Len_{\text{task}_i}$  denotes the length of the task output, and  $Len_{\text{total}}$  is the total sequence length. And the length measurement is conducted at the token level.}

\paragraph{\textbf{Training Process.}}
During training, we consider that the audio in our heterogeneous datasets comes from different sources, leading to noticeable differences in audio features. To help UniSLU adapt to these variations, we fully fine-tune both the encoder and decoder. This allows the model to learn better across diverse audio characteristics, improving its overall performance on both datasets.
\section{Experiments}
\paragraph{\textbf{Datasets.}} Following recent works \cite{DBLP:conf/asru/DenisovV23, DBLP:journals/corr/abs-2409-08107}, we use SLUE-VoxPopuli and SLUE-VoxCeleb \cite{shon2022slue} as two heterogeneous and open-source datasets \footnote{https://huggingface.co/datasets/asapp/slue} for spoken NER and spoken SA tasks, which contain speech-text aligned datasets with NER or SA annotations, respectively.
SLUE-VoxPopuli is an English dataset sourced from the European Parliament event recording datasets VoxPopuli \cite{wang2021voxpopuli}, containing audio/transcript paired sentences and annotated named entities within the sentences. The entity types include PLACE, QUANT, ORG, WHEN, NORP, PERSON, and LAW. SLUE-VoxCeleb is based on English single-speaker YouTube conversations, with audio/transcript paired sentences and the annotated sentiment label (i.e., positive, negative, neutral, and mixed) for each sentence.
Table \ref{tab:dataset_sizes} shows the detailed statistics of the two datasets. 
\begin{table}[t]
    \setlength{\tabcolsep}{3pt}
    \centering
    \caption{Statistics of SLUE-VoxPopuli and SLUE-VoxCeleb, including the number of the sentences (\#Sent), the number of the entities (\#NE) and the time duration of the audios (Dur).}
    \begin{tabular}{llllll}
        \toprule
        \textbf{Datasets}      &  & \textbf{Train} & \textbf{Dev} & \textbf{Test} & \textbf{Tasks}\\ 
        \midrule
        \multirow{3}{*}{\shortstack{SLUE \\ VoxPopuli}}         & \#Sent & 5,000 & 1,753 & 1,842 & \multirow{3}{*}{\shortstack{ASR \\ NER}}  \\
                               & \#NE  & 5,730 & 1,819 & 1,961 & \\
                               & Dur  & 14.5h & 5.0h & 4.9h & \\
        \midrule
        \multirow{2}{*}{\shortstack{SLUE \\ VoxCeleb}}          & \#Sent & 5,777        & 1,454  & 3,553 & \multirow{2}{*}{\shortstack{ASR \\ SA}}  \\ 
                               & Dur  & 12.8h  & 3.2h & 7.8h & \\
        \bottomrule
    \end{tabular}
    \label{tab:dataset_sizes}
\end{table}
\begin{table}[h!]
\setlength{\tabcolsep}{1pt}
\centering
\caption{Hyper-parameters and model details used in pre-training and fine-tuning.}
\begin{tabular}{l l|l l}
    \hline
    \multicolumn{2}{c}{\textbf{Whisper-small}} & \multicolumn{2}{c}{\textbf{Whisper-medium}} \\
    \hline
    Parameters & 244M & Parameters & 769M \\
    Encoder Blocks & 12 & Encoder Blocks & 24 \\
    Decoder Blocks & 12 & Decoder Blocks & 24 \\
    Hidden Size & 768 & Hidden Size & 1024 \\
    Fine-tune Epoch & 100 & Fine-tune Epoch & 100 \\
    Learning Rate & 1e-5 & Learning Rate & 2e-5 \\
    Warm Ratio & 0.01 & Warm Ratio & 0.01 \\
    Log Mel & 80 & Log Mel & 80 \\
    Window Size & 25ms & Window Size & 25ms \\
    Window Shift & 10ms & Window Shift & 10ms \\
    Num-beams & 5 & Num-beams & 5 \\
    \hline
    \multicolumn{2}{c}{\textbf{Qwen2.5-3B-Instruct}} & \multicolumn{2}{c}{\textbf{Qwen2.5-7B-Instruct}} \\
    \hline
    Layers & 36 & Layers & 28 \\
    Heads(KV) & 16/2 & Heads(KV) & 28/4 \\
    Pretrain Epoch & 2 & Pretrain Epoch & 2 \\
    Fine-tune Epoch & 50 & Fine-tune Epoch & 50 \\
    Learning Rate & 2e-4 & Learning Rate & 2e-4 \\
    Lora-r & 32 & Lora-r & 32 \\
    Lora-alpha & 32 & Lora-alpha & 32 \\
    Lora-dropout & 0.05 & Lora-dropout & 0.05 \\
    Temperature & 0.7 & Temperature & 0.7 \\
    Top-p & 0.9 & Top-p & 0.9 \\
    Repetition-penalty & 1.2 & Repetition-penalty & 1.2 \\
    Num-beams & 1 & Num-beams & 1 \\
    \hline
\end{tabular}
\label{tab:hyperparameters}
\end{table}
\paragraph{\textbf{Implemention Detail.}}

Our base model uses Whisper-small \footnote{https://huggingface.co/openai/whisper-small}, with a 12-layer Transformer block for the encoder and a 12-layer decoder. The model is fine-tuned for 100 epochs with a learning rate of 1e-5 and a batch size of 50. {\changed Additionally, we also conduct experiments on Whisper-medium \footnote{https://huggingface.co/openai/whisper-medium}, which adopts a larger encoder-decoder architecture, with both the encoder and decoder comprising 24 transformer blocks. We run each of our models three times with different random seeds and report the average result. For fine-tuning, our smaller unified model requires approximately 14 hours on V100-32G GPUs, while the larger unified model takes around 20 hours on A100-40G GPUs. Both model sets use Whisper’s tokenizer and we provide more detailed parameter settings in the Table \ref{tab:hyperparameters}.\par}

\begin{table*}[t]
    \setlength{\tabcolsep}{10pt}
    \centering
     \caption{Performance of the proposed UniSLU model and other compared models across ASR, NER, and SA tasks, along with the overall SLUE SCORE on the test set. VP and VC denote the SLUE-VoxPopuli and SLUE-VoxCeleb datasets, respectively. Models marked with $\dagger$ indicate the results reported in SLUE Benchmark \cite{shon2022slue}.}
    \begin{tabular}{lccccc}
    \toprule
    \multirow{2}{*}{\textbf{Model}} & \multicolumn{2}{c}{\textbf{ASR-WER}} & \textbf{NER-F1} & \textbf{SA-F1} & \multirow{2}{*}{\textbf{SLUE SCORE}} \\
                                    & \textbf{VP} & \textbf{VC}                  & \textbf{VP}          & \textbf{VC}          &                                   \\
    \midrule
    \multicolumn{6}{c}{\cellcolor{gray!25} Separate Models} \\
    $\text{SLUE (W2V2-B-LS960)}^{\dagger}$      & 18.40  & 20.90  & 49.60  & 48.60  & 59.40        \\
    $\text{SLUE (W2V2-L-LL60K)}^{\dagger}$      & 12.10  & 13.80  & 50.50  & 50.10  & 62.55   \\
    $\text{SLUE}$ (Whisper-small)           & 9.80 & 13.20 & 61.89 & 51.91 & 67.43    \\
    $\text{SLUE}$ (Whisper-medium)       & 8.21 & \textbf{10.62} & 57.18 & 52.58 & 66.78 \\
    GenSLU (Whisper-small)                   & 10.20 & 12.54 & 50.29 & 59.69 & 66.17   \\
    GenSLU (Whisper-medium)                  & 8.63  & 10.92 & 63.83 & 60.92 & 71.65        \\
    \midrule
    \multicolumn{6}{c}{\cellcolor{gray!25} Unified Models}  \\
    UniSLU (Whisper-small)             & 10.58 & 13.10 & 59.61 & 59.14 & 68.97   \\
    \phantom{xxx}w/o Dynamic Loss                         & 10.04 & 12.85 & 51.24 & 59.94 & 66.57   \\
    \phantom{xxx}w/o Fine-tuning Encoder                  & 11.13 & 13.52 & 47.57 & 57.24 & 64.16   \\
    \hdashline
    UniSLU (Whisper-medium)            & 8.77 & 11.47 & \textbf{69.95} & 61.60 & \textbf{73.81} \\
    \phantom{xxx}w/o Dynamic Loss             & 8.97 & 10.93 & 68.69 & \textbf{61.74} & 73.49 \\
    \phantom{xxx}w/o Fine-tuning Encoder      & \textbf{8.08} & 11.05 & 65.18 & 61.22 & 72.27 \\

    \bottomrule
    \end{tabular}
    \label{tab:slue_results_test}
\end{table*}
\paragraph{\textbf{Evaluation Metrics.}} 
For ASR, we adopt the widely used Word Error Rate (WER) \cite{omachi-etal-2021-end} as our metric. For NER, following previous works \cite{ghannay2018end, yadav2020end}, we use the Micro-F1 score to evaluate the performance regarding both entity boundaries and entity types. For SA, in line with previous works \cite{bertero2016real, lu2020speech}, we use the Macro-F1 score to measure the performance.
Besides the above task-specific metrics for ASR, NER, and SA, we apply the SLUE SCORE metric \cite{shon2022slue} to evaluate the overall SLU performance on the three tasks, which is defined as: 
$SLUE_{score}=\frac{1}{3}((100-\frac{WER_{VP}+WER_{VC}}{2})+F1_{VP}+F1_{VC})$
where  $\text{WER}_{\text{VP}}$ and $\text{WER}_\text{VC}$ represent the WER score of the VP and VC datasets,  $\text{F1}_\text{VP}$  denotes the Micro-F1 score of the NER performance on VP, and  $\text{F1}_\text{VC}$  is the Macro-F1 score of SA on VC. 


\subsection{Main Results} 
\label{sec:main-results}
The main results on the SLUE-VoxPopuli and SLUE-VoxCeleb test sets are shown in Table \ref{tab:slue_results_test}. 
``SLUE (W2V2-B-LS960)'' and ``SLUE (W2V2-L-LL60K)'' represent the methods from \citet{shon2022slue}, which train spoken NER and SA using separate models. In these methods, spoken NER is treated as an ASR task, with specific symbols used to denote named entities (NEs), while spoken SA is handled with an additional classification module following the audio encoder. These methods use Wav2Vec 2.0 \cite{baevski2020wav2vec} as the audio encoder, with versions W2V2-B-LS960 and W2V2-L-LL60K, respectively. {\changed ``$\text{SLUE}$ (Whisper-small)'' refers to a method similar to SLUE, where Wav2Vec 2.0 is replaced with Whisper. For both NER and SA, we follow the same formulations as in SLUE. Specifically, for NER, we use a special tagging format, while for SA, we integrate a classification module after the Whisper encoder, which includes a self-attention pooling layer and two fully connected layers, including the output layer.} ``GenSLU'' refers to our generative method that performs both spoken NER and SA tasks by generating the corresponding NE or SA outputs as suffixes to the transcript, with the two tasks trained separately. ``UniSLU'' is our proposed method as illustrated in Section \ref{sec:approach}, with versions whisper-small and whisper-medium respectively.
\paragraph{\textbf{Results of Separate Models.}} 
{\changed To compare with unified models and better understand the impact of model backbones, task formulations, and parameter scales on SLU performance, we first report the results of different separate models in the separate model part of Table~\ref{tab:slue_results_test}.

First, we compare the first four rows in this part, where the only difference lies in the backbone model—replacing W2V2 with Whisper—while keeping the task-specific formulations the same (i.e., NER with special tags and SA as a classification task). The results show that ``$\text{SLUE}$ (Whisper-small)'' consistently outperforms ``SLUE (W2V2-L-LL60K)'', demonstrating that the robust Whisper encoder provides a stronger foundation for downstream SLU tasks such as spoken NER and SA.

Second, when comparing ``GenSLU (Whisper-small)'', which adopts a fully generative formulation for ASR, NER, and SA, with ``SLUE(Whisper-small)'', we observe a considerable performance boost, particularly in the SA task. This improvement stems primarily from the generative approach’s ability. Unlike traditional classification methods that treat labels as independent categories, the generative method naturally connects sentiment labels to the surrounding textual context, enabling richer semantic understanding and leading to better SA performance. For NER, however, the results suggest that the special tagging strategy remains more effective in small models, as it allows entities to be recognized immediately upon occurrence, reducing issues such as forgetting in longer sequences.

Third, we analyze the performance gap between the SLUE methods and generative methods when using the larger Whisper-medium model. While the generative method continues to deliver strong SA results, the tag-specific NER approach does not show a similar improvement. We believe this is because tagging them inline offers limited gains from increased model capacity. Moreover, as model size increases, the divergence between the tagging format and the original pretraining objective becomes more pronounced, making it harder for the model to converge effectively. As a result, the NER performance of the tag-specific method on Whisper-medium can even degrade compared to its performance on Whisper-small.

Finally, we examine the generative approach across different model sizes. The results show that ``GenSLU (Whisper-medium)'' achieves further performance gains compared to ``GenSLU (Whisper-small)'', indicating that the generative modeling method can better utilize the enhanced semantic capacity of larger models. This scalability highlights the advantage of using a generative formulation for unified SLU, especially in complex multimedia scenarios.}
\begin{figure*}[t]
  \centering
  \includegraphics[width=\textwidth]{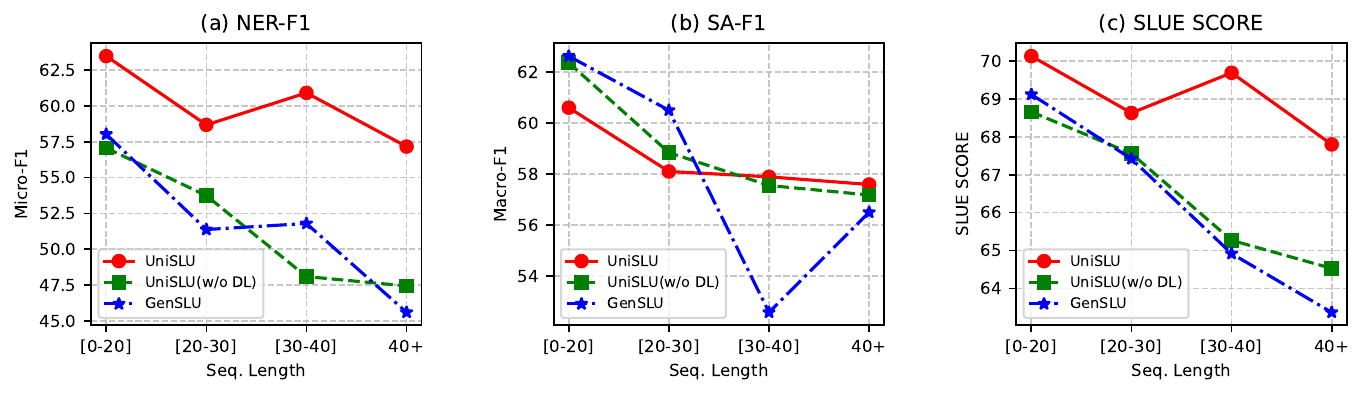} 
  \caption{Comparison of Micro-F1, Macro-F1 and SLUE SCORE metrics across different sequence length categories on test set. The sequence lengths on the horizontal axis are measured in words.}
  \label{fig:seq_len}
\end{figure*}
\paragraph{\textbf{Results of our Unified Model.}} 
{\changed We further evaluate the effectiveness of our proposed unified generative framework, UniSLU, by comparing it with separate models across the three main SLU tasks: ASR, NER, and SA. The results reveal several key observations.

First, compared to separate models, ``UniSLU (Whisper-small)'' achieves higher overall performance, particularly with clear improvements in the NER task. For instance, when comparing ``$\text{SLUE}$ (Whisper-small)'' with ``UniSLU (Whisper-small)'', we observe that the unified model maintains strong SA performance while further improving NER relative to ``GenSLU (Whisper-small)''. Although it still slightly underperforms ``$\text{SLUE}$ (Whisper-small)'' in NER alone, the overall SLUE SCORE is superior. This improvement can be attributed to UniSLU's ability to jointly train on both SLUE-VoxPopuli and SLUE-VoxCeleb, whereas the separate SLUE models use only VoxPopuli for NER. This broader data coverage provides more diverse and informative entity examples. Moreover, the unified framework enables multitask interaction, allowing the model to share semantic representations across ASR, NER, and SA, which in turn improves its ability to understand contextual relationships and recognize entities more accurately.

Second, when scaling up to ``UniSLU (Whisper-medium)'', we observe consistent performance improvements across all three SLU tasks. Compared to ``UniSLU (Whisper-small)'', the medium-sized model benefits from a larger encoder and deeper decoder, resulting in stronger semantic representation and sequence modeling capabilities. The enhanced encoder leads to more accurate ASR transcriptions, providing cleaner textual inputs for downstream tasks, while the deeper decoder helps reduce forgetting in longer sequences—particularly beneficial for the NER task. These architectural improvements, combined with the unified generative formulation, further enhance the model’s capacity for effective multitask learning and semantic understanding.

In summary, separate models benefit from the strong Whisper backbone and the generative method but lack cross-task interaction and shared learning. Our unified framework, UniSLU, outperforms these models by leveraging heterogeneous data and enabling unified modeling across tasks. With larger model capacity, UniSLU (Whisper-medium) achieves further gains. These results highlight the effectiveness and scalability of our generative approach, particularly in complex, multimodal SLU scenarios.}
\subsection{Ablation Study}
To better understand the impact of key design choices in our model, we conduct an ablation study focusing on two components: the dynamic loss mechanism and encoder fine-tuning. These elements are critical for balancing multi-task learning and improving representation quality. The results, shown in Table~\ref{tab:slue_results_test}, highlight their importance in achieving strong SLU performance.
\paragraph{\textbf{Impact of Dynamic Loss.}}
The ``w/o Dynamic Loss'' row in the lower part of every unified model in Table \ref{tab:slue_results_test} refers to the results obtained from the UniSLU model fine-tuned using the original loss function, without any modifications. 
{\changed In the ``UniSLU (Whisper-small)'' part, the comparison between UniSLU and ``w/o Dynamic Loss'' highlights that introducing dynamic loss significantly enhances overall performance—especially in NER-F1. This improvement comes from the fact that, in most datasets, the ASR outputs are typically much longer than those of the other SLU tasks, such as NER and SA, as shown in Figure~\ref{fig:data_representation}. When a standard cross-entropy loss—specifically, without the use of the dynamic loss—is applied across the entire output sequence it tends to bias training toward the ASR task. If left unaddressed, this imbalance can lead to underfitting in NER and SA due to insufficient optimization focus, and potential overfitting in ASR, ultimately resulting in overall performance degradation.
To address this imbalance, we introduce a dynamic loss mechanism that adjusts task-specific loss weights based on the length of each task’s output sequence at the token level. This ensures that no single task dominates the training process, allowing the model to learn all tasks more evenly. Without this adjustment, shorter-output tasks like NER and SA may receive insufficient focus, which can lead to weaker performance.
In the ``UniSLU (Whisper-medium)'' section, the comparison with ``w/o Dynamic Loss'' reveals that increasing the number of encoder-decoder layers enhances the model’s ability to manage multiple tasks. The larger model is more capable of handling long sequences, balancing shared and task-specific knowledge, and fostering better cross-task interactions. These enhancements raise the performance ceiling of UniSLU, making the benefits of dynamic loss relatively less pronounced as the model size grows.

Overall, dynamic loss proves especially valuable in low-resource settings and when using smaller models, where it provides a notable boost in performance and promotes more balanced learning across all tasks.}

\paragraph{\textbf{Fine-tuned vs. Frozen Whisper Encoder.}}
We investigate the impact of fine-tuning versus freezing the Whisper encoder.
{\changed In the ``UniSLU (Whisper-small)'' section, comparing ``UniSLU'', which uses a fine-tuned Whisper encoder, with the ``w/o Fine-tuning Encoder'' model that keeps the encoder frozen, we observe that fine-tuning the encoder improves performance across ASR, NER, and SA.
However, the larger encoder in the Whisper-medium model can significantly mitigate the negative impact of freezing, especially in ASR. In fact, due to its strong capability in extracting audio information, the frozen encoder sometimes even achieves better ASR results. This improved ASR performance also helps reduce the gap in SLU performance between the frozen and fine-tuned encoder settings.
Therefore, for smaller models, full multitask fine-tuning is a better choice. Interestingly, a closer look at the model configurations in Table \ref{tab:hyperparameters} reveals that Whisper-medium has exactly twice the number of transformer blocks as Whisper-small. As a result, fully fine-tuning Whisper-small and fine-tuning only the decoder of Whisper-medium (with approximately 380M trainable parameters) yield comparable trainable parameter counts. This suggests that in low-resource scenarios—particularly in real-world multimedia settings—freezing the encoder of a larger model can maximize performance while significantly reducing computational costs, potentially even outperforming the full fine-tuning of a smaller model.

Overall, fine-tuning the encoder brings more consistent benefits in smaller models, while larger models show greater robustness to encoder freezing.}
\begin{table*}[t]
    \centering
    \setlength{\tabcolsep}{5pt}
    \caption{Comparison of performance metrics for LLMs across WER (VP/VC), NER (F1), and SA (F1) on the test set. The model marked with $\dagger$ denote results originally reported in the SLUE Benchmark \cite{shon2022slue}.}
    \begin{tabular}{lccccc}
    \toprule
    \multirow{2}{*}{\textbf{Model}} & \multicolumn{2}{c}{\textbf{ASR-WER}} & \textbf{NER-F1} & \textbf{SA-F1} & \multirow{2}{*}{\textbf{SLUE SCORE}} \\
                                    & \textbf{VP} & \textbf{VC}                  & \textbf{VP}          & \textbf{VC}          &                                   \\
    \midrule
    \multicolumn{6}{c}{\cellcolor{gray!25} w/o LLM Decoder}  \\
    $\text{SLUE (W2V2-B-LS960)}^{\dagger}$     & 18.40  & 20.90  & 49.60  & 48.60  & 59.40        \\
    UniSLU (Whisper-small)        & \textbf{10.58} & \textbf{13.10} & 59.61 & 59.14 & 68.97        \\
    \multicolumn{6}{c}{\cellcolor{gray!25} LLM Decoder}  \\
    UniSLU (W2V-BERT-2.0 + Qwen2.5-3B)      & 16.14 & 23.24 & 60.24 & 60.73 & 67.09        \\
    UniSLU (Whisper-small + Qwen2.5-3B)     & 14.74 & 19.01 & 61.57 & 63.36 & 69.35        \\
    UniSLU (Whisper-small + Qwen2.5-7B)     & 12.20 & 16.40 & \textbf{67.79} & \textbf{64.76} & \textbf{72.75} \\
    \bottomrule
    \end{tabular}
    \label{tab:performance_metrics}
\end{table*}
\subsection{Sequence Length Analysis}
From Figure~\ref{fig:seq_len}, we conduct a detailed analysis of the performance of three representative generative models on samples with varying sequence lengths. Our evaluation focuses on three key aspects of SLU: NER, SA, and overall task performance as reflected by the SLUE SCORE. This analysis aims to reveal how model robustness and task effectiveness are influenced by input length, which is a common challenge in real-world spoken content, especially in multimedia applications where utterance lengths can vary significantly.

Starting with the left plot (a), we observe that the NER performance of all three models declines to different extents as the sequence length increases. However, UniSLU consistently outperforms the others across all length ranges and shows a more gradual performance drop. This suggests that our dynamic loss mechanism effectively balances the training process by assigning a more appropriate weight to the NER task, thereby reducing the impact of longer sequences.

In the middle plot (b), while SA performance also declines with increasing length, the unified model (UniSLU) demonstrates greater stability compared to the separate model, which exhibits more noticeable variability in the 30–40 word range. We attribute this instability to the limited number of samples in the SLUE-VoxCeleb dataset within that range, which leads to weaker semantic representation. In contrast, UniSLU benefits from the inclusion of SLUE-VoxPopuli data, and through multitask learning, it compensates for the samples in this segment. This cross-task interaction enhances the model’s understanding and enables it to maintain stronger performance.

Lastly, in the right plot (c), the overall SLUE SCORE follows a trend similar to NER, with performance gradually declining as the sequence length increases. Nevertheless, UniSLU consistently achieves better results by fully leveraging heterogeneous data sources and reinforcing interaction between tasks. Furthermore, the use of dynamic loss makes the model more robust to challenges posed by longer sequences, leading to stronger and more balanced overall performance.

\subsection{Combining LLMs} \label{LLM}

As outlined in Section~\ref{framework}, our unified model framework can also be easily extended to integrate with LLMs. {\changed In our experiments that combined with LLM decoder, we use W2V-BERT-2.0 \footnote{https://huggingface.co/facebook/w2v-bert-2.0} and Whisper-Encoder as audio encoders, Qwen2.5-3B \footnote{https://huggingface.co/Qwen/Qwen2.5-3B-Instruct} \cite{yang2024qwen2} and Qwen2.5-7B-Instruct \footnote{https://huggingface.co/Qwen/Qwen2.5-7B-Instruct} as the LLM decoders. To connect the audio encoder with the subsequent LLM decoders, we design an adapter consisting of two linear layers. The audio features output by the audio encoder pass through this adapter before being input into the LLM. Given the adaptation challenge between the audio encoder and the LLM, we pre-train the adapter on LibriSpeech-960 (LS960) before fine-tuning it.
During pre-training, the model is trained for 2 epochs with a learning rate of 2e-4 and a warm-up ratio of 0.01, while updating only the parameters of the adapter module. In the fine-tuning phase, the model is trained for 50 epochs using the same learning rate and warm-up ratio, with LoRA applied to update the Q, K, V, and O projection layers in transformer blocks. Pre-training and fine-tuning on the 7B model take approximately 9 hours and 12 hours, respectively, and are conducted on two A100-40G GPUs. Further details on the training parameters are provided in Table~\ref{tab:hyperparameters}.

In Table~\ref{tab:performance_metrics}, we compare the ``SLUE (W2V2-B-LS960)'' model and ``UniSLU (Whisper-small)'' with three sets of large-model experiments. Among these five experiments, only ``UniSLU (Whisper-small)'' is not pre-trained on LS960. This decision is based on the observation that Whisper-small achieves a WER of 8.99 on the LibriSpeech test set without any additional training. We believe this is likely because the LS960 dataset is already included in the extensive pre-training data of Whisper-small, suggesting that additional pre-training on LS960 may not be necessary. Meanwhile, ``SLUE (W2V2-B-LS960)'' serves as a reference model that is pre-trained on LS960.

First, by comparing ``UniSLU (W2V-BERT-2.0 + Qwen2.5-3B)'', ``UniSLU (Whisper-small)'', and ``UniSLU (Whisper-small + Qwen2.5-3B)'', we observe a clear improvement in overall SLU performance (SLUE SCORE) after incorporating the LLM. While ASR performance slightly declines, both NER and SA show notable gains. We attribute the ASR drop to limited pretraining of the LLM-based decoder compared to Whisper’s original decoder, as well as incomplete adaptation between the audio encoder and LLM. Despite this, the LLM significantly enhances NER and SA by contributing stronger semantic understanding.

Second, comparing ``UniSLU (Whisper-small + Qwen2.5-3B)'' and ``UniSLU (Whisper-small + Qwen2.5-7B)'' shows that larger LLMs further boost performance across all tasks. This demonstrates that our generative and unified modeling approach is well-positioned to benefit from increased model capacity.}
In summary, integrating LLMs into our unified framework significantly boosts NER and SA by enhancing semantic understanding, though ASR gains remain limited. These results further support the effectiveness of unified modeling, especially in complex multimedia scenarios.
\section{Conclusion}

In this work, we propose UniSLU, a unified generative framework that facilitates diverse SLU tasks within a single model architecture, enhances cross-task interactions, and fully leverages heterogeneous datasets across tasks. To make full use of varied available heterogenous datasets across tasks, we introduce a unified representation that integrates diverse SLU tasks into a consistent format, allowing comprehensive utilization of heterogeneous data. 
Additionally, our framework unifies ASR, spoken NER, and SA within a single generative model, fostering deeper task interaction and enabling seamless integration with LLMs to leverage their strong generative capabilities. Extensive experiments conducted on public datasets demonstrate the effectiveness of our approach, with UniSLU consistently outperforming several strong baselines. These results validate the potential of UniSLU to significantly improve SLU performance and lay a foundation for future research on unified SLU systems. While our current work focuses on three representative SLU tasks and primarily English data, future work will extend to a broader range of tasks, languages, and application domains.

\bibliography{custom}
\end{document}